# DWARF GALAXIES MIGHT NOT BE THE BIRTH SITES OF SUPERMASSIVE BLACK HOLES

Intermediate-mass black holes (BHs) in local dwarf galaxies are considered the relics of the early seed BHs. However, their growth might have been impacted by galaxy mergers and BH feedback so that they cannot be treated as tracers of the early seed BH population.
**Mar Mezcua**

The discovery that quasars with black hole (BH) masses of more than $10^9$ $M_{Sun}$ already existed ~700 Myr after the Big Bang has been puzzling the astronomical community for more than a decade[1]. How did such massive BHs have time to form in such a young Universe? A wealth of theoretical models propose that they grew from lower-mass seed BHs of ~100-$10^5$ $M_{Sun}$ via accretion and mergers[2]. Such seed BHs could have formed from the death of the first generation of (population III) stars, from direct collapse of pristine gas, or from mergers in dense stellar clusters. Detecting these early seeds is extremely challenging with current instrumentation; however, those that did not become supermassive should be found in the local Universe as leftover intermediate-mass BHs (IMBHs) of 100-$10^5$ $M_{Sun}$ (see Fig. 1).

It is often presumed that dwarf galaxies have not significantly grown through merger and accretion and are very likely to resemble the primordial galaxies of the infant Universe[3]. Simulations predict that a large fraction of today's dwarf galaxies should host 'light' (100-1,000 $M_{Sun}$) IMBHs if seed BHs formed from population III stars, while a lower occupation fraction of 'heavy' ($10^4$-$10^5$ $M_{Sun}$) BHs is expected if seed BHs formed from direct gas collapse[3]. Thus, deriving the BH occupation fraction in dwarf galaxies seems to be pivotal for understanding how the early seed BHs formed and evolved into the supermassive BHs observed today.

Several tens of actively-accreting IMBH candidates have been found in dwarf galaxies as low-mass active galactic nuclei (AGNs) with BH masses ≤$10^6$ $M_{Sun}$ [4,5]. This has allowed us to derive an AGN fraction that, when taken as an upper limit to the true BH occupation fraction, seems to favour the direct collapse scenario for the formation of BH seeds[6]. This scenario is supported by the finding that the correlation between BH mass and stellar velocity dispersion of massive galaxies flattens in the low-mass regime ($10^5$-$10^6$ $M_{Sun}$ [7,8]), as predicted by formation models of direct collapse seed BHs. Despite these presumably significant advances in the field, no studies have yet proven whether these tens of low-mass AGNs are the relics of the early seed BHs.

Several caveats seem to be, at least observationally, overlooked. First of all, dwarf galaxies undergo on average three major mergers after their formation[9], and thus have a merger frequency much higher than that of massive galaxies. Dwarf galaxy mergers may lead to the coalescence of their central BHs and could trigger rapid BH accretion, both of which processes would lead to significant growth of the primordial seed[10,11] (Fig. 1). In this case, the low-mass AGNs in dwarf galaxies should not be treated as relics of the early Universe seed BHs, which has important implications for distinguishing between seed BH formation scenarios using local dwarf galaxies. Having dwarf galaxies whose BHs have grown efficiently through galaxy mergers would, for instance explain why only 'heavy' BHs of ~$10^5$ $M_{Sun}$ are detected in local dwarf galaxies even though 'light' seed BHs are predicted to be more abundant. The latter has been often attributed to an observational bias of optical IMBH searches, which are skewed toward high Eddington ratios and high luminosities[4,12,13]. Such bias could be overcome by increasing the low number of sub-Eddington accreting low-mass AGN so far detected[6,14,15] via a systematic search of IMBHs in the radio regime. Evidence is growing that both major mergers of dwarf galaxies and minor mergers could trigger AGN activity[15-18]. A thorough investigation of the role of mergers in growing BHs in dwarf galaxies and its implication for understanding whether these BHs are of primordial origin is thus urgent.

A second aspect challenging seed BH studies by means of local dwarf galaxies is AGN feedback. Most numerical simulations indicate that outflows from young stars and supernovae evacuate gas from the nucleus of dwarf galaxies, stunting BH growth and thus the impact of AGN feedback[19,20]. However, others indicate that AGN feedback has the biggest impact on dwarf galaxies[21,22]. Observationally the results are also controversial: Martin-Navarro & Mezcua [8] find observational support for the supernova feedback scenario, while Penny et al.[23] report that AGN feedback could impact star formation in dwarf galaxies. 'Positive' AGN feedback can trigger star formation (see, for example, [24]) in turn impacting the fodder for the BH. In dwarf galaxies hosting AGNs, BH growth might thus not be hampered by supernova feedback but enhanced by AGN feedback. Clarifying the role that stellar and AGN feedback processes play in regulating BH growth in dwarf galaxies is necessary before we continue assuming that due to supernova feedback neither the galaxy nor the BH have grown much over cosmic time.

If mergers and AGN feedback hamper the use of IMBHs in dwarf galaxies as tracers of the early Universe seed BHs, how can we probe the existence of the high-redshift supermassive BH progenitors? One possibility could be to search for IMBHs in extremely metal-poor galaxies. These relatively bright, low-mass galaxies with a fraction of metals below one-tenth that of the Sun are in a current phase of intense star formation similar to high-redshift galaxies[25]. The majority of extremely metal-poor galaxies do not show signs of recent merger events and they generally reside in low density environments[26]. The regions of star formation in the galaxy outskirts exhibit a metallicity drop[27,28] and could harbor pockets of pop III stars[29], suggesting that accretion of metal-poor gas via cold flows may trigger and sustain the star formation in these galaxies[28]. These properties convincingly demonstrate that extremely metal-poor galaxies are the best local analogs of the first galaxies and that they should be used as probes of the primordial conditions of the Universe, namely the detection of seed BHs. Yet, few IMBH studies have focused on extremely metal-poor galaxies aimed at probing whether they host the early Universe seed BHs[30,31].

The next generation of major facilities (for example the *James Webb Space Telescope*, the *Athena X-ray Observatory* or the *Lynx X-ray Observatory* mission concept) could provide direct detection of the early seeds, which is one of the key science cases planned for these observatories.

The detection of gravitational wave emission from an IMBH merger will provide irrefutable proof that IMBHs exist. While the coalescence of BHs with masses greater than 1,000 $M_{Sun}$ will produce gravitational waves with frequencies only detectable by space-based interferometers, the ring-down of binaries with total mass ~200 - 2,000 $M_{Sun}$ should produce a signal detectable by the advanced LIGO and Virgo interferometers[32,33].

Only once we understand the role of mergers and feedback processes in dwarf galaxies will we be able to understand whether low-mass AGNs are robust probes of the early BH seeds. The question of how supermassive BHs form is still open, and will probably remain so for at least another decade.

*Mar Mezcua[1,2]*
[1]*Institute of Space Sciences (ICE, CSIC), Barcelona, Spain.* [2]*Institut d'Estudis Espacials de Catalunya (IEEC), Barcelona, Spain.*
*e-mail:* [marmezcua.astro@gmail.com](marmezcua.astro@gmail.com)

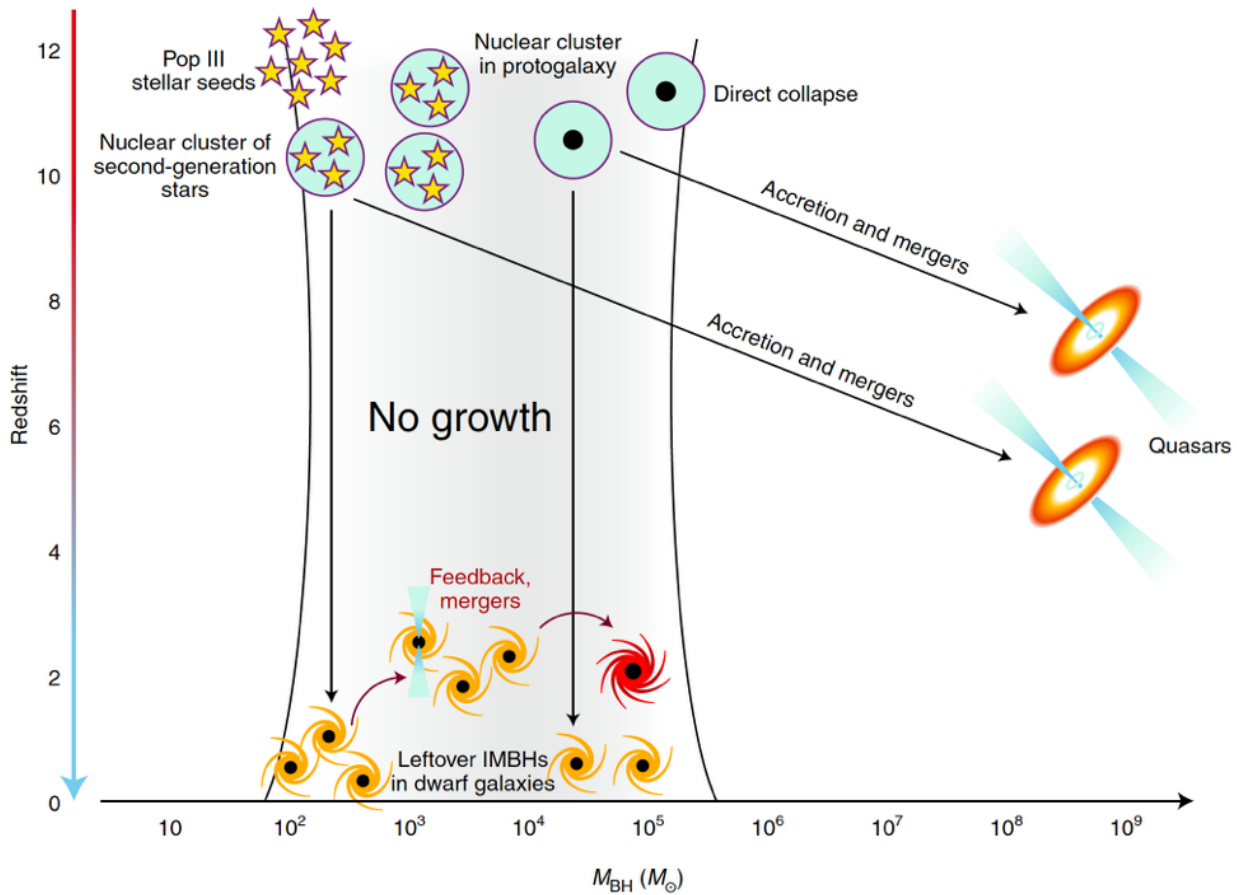

**Fig. 1 - Formation and evolution of seed black holes.** Seed BHs in the early Universe could form from population III stars, from mergers in dense stellar clusters (formed out either from the second generation of stars or from inflows in protogalaxies), or from direct collapse of dense gas in protogalaxies, and grow via accretion and merging to $10^9$ $M_{Sun}$ by redshift~7. Those seed BHs that did not become supermassive should be found in the local Universe as leftover IMBHs in dwarf galaxies. However, dwarf galaxy mergers and BH feedback could significantly impact the growth of these leftover seeds. In this way, for instance, some of the pop III remnants could become 'heavy' BHs ($M_{BH}$ ~$10^5$ $M_{Sun}$; galaxy marked as red), in which case local IMBHs should not be considered relics of the early Universe seed BHs. Credit: adapted from ref.[7], World Scientific.